\documentstyle[aps,prl,preprint]{revtex} 
\begin{document} 
\bibliographystyle{unsrt} 
 
\draft 
\title{Discrete breathers in systems with homogeneous potentials - 
analytic solutions} 
\author{A. A. Ovchinnikov$^1$ and S. Flach} 
\address{Max-Planck-Institute for the Physics of Complex Systems, 
N\"othnitzer Str. 38, D-01187 Dresden, Germany  
} 
\date{\today} 
\maketitle 
\begin{abstract} 
We construct lattice Hamiltonians with homogeneous interaction potentials 
which allow for explicit breather solutions. Especially 
we obtain exponentially localized solutions for 
$d$-dimensional lattices with $d=2,3$. 
\end{abstract} 
 

\section{Introduction} 
 
The understanding of dynamical localization in classical spatially extended 
and ordered systems \cite{aao70} experienced recent considerable progress 
\cite{sa97}, \cite{sfcrw98}, \cite{physicad119}. Specifically 
time-periodic and spatially localized solutions of the classical equations 
of motion exist, which are called (discrete) breathers, or intrinsic 
localized modes. The attribute discrete stands for the spatial discreteness of 
the system, i.e. instead of field equations one typically considers the 
dynamics of degrees of freedom ordered on a spatial lattice.  
The lattice 
Hamiltonians are invariant under discrete translations in space. 
The discreteness 
of the system produces a cutoff in the wavelength of extended states, and 
thus yields a finite upper bound on the spectrum of eigenfrequencies $\Omega_q$ 
(phonon band) of 
small-amplitude plane waves (one usually 
assumes that usually for small amplitudes  
the Hamiltonian is in leading order a quadratic form of the degrees of  
freedom). If now the equations of motion contain nonlinear terms, the 
nonlinearity will in general allow to tune frequencies of periodic orbits 
outside of the phonon band, and if all multiples of a given frequency 
are outside the phonon band too, there seems to be no further barrier 
preventing spatial localization (for reviews see \cite{sa97},\cite{sfcrw98}). 

Discrete breathers have been recently experimentally detected in
weakly coupled waveguides \cite{hseysrmarbjsa98}, MX solids 
\cite{sblssbws99} and Josephson junction ladders \cite{pbdaavusfyz99}.
The broad spectrum of applicability of the localization concept
described above makes it worthwhile to continue theoretical efforts
to characterize the properties of discrete breathers.
 
So far there is very little knowledge about explicit forms of 
breather solutions. Except for trivial limiting cases like the 
antiintegrable limit, i.e. the case of uncoupled degrees of freedom 
\cite{ma94}, we know only about the solutions 
of the Ablowitz-Ladik lattice \cite{al76} (an integrable  
one-dimensional variant of the nonlinear Schr\"odinger equation). 
What is generically available is the abstract knowledge 
about existence or nonexistence of discrete breathers for a specific 
system, and the spatial decay properties far from the center of 
the breather, where due to the smallness of the amplitudes linearizations 
or other perturbation techniques are applicable. 
Note that generically breathers can appear in nonintegrable systems. 
 
From the above it appears that nonintegrability spoils the possibility 
to obtain analytic forms of the breather solution. We will show 
that this is not the case by constructing Hamiltonians which allow 
for explicit solutions and are most probably not integrable.  
Moreover we will even construct solutions for two- and three-dimensional 
lattices. Although these models are not motivated by certain 
applications, the study of their properties can be helpful 
with respect to discrete breathers. 
 
\section{A bond-ordered quasi-linear chain} 
 
In this section we consider a one-dimensional model with 
the Hamiltonian 
\begin{equation} 
H = W_k + W_p\;\;,\;\;W_k = \sum_l \frac{1}{2}p_l^2 \;\;, 
\;\; W_p= \frac{1}{2} \sum_l (x_l - x_{l+1})^2 h(s_l)\;\;,\;\; 
s_l(\{x_{l'}\}) = \frac{x_l}{x_{l+1}} + \frac{x_{l+1}}{x_l} 
\;\;. 
\label{2-1} 
\end{equation} 
The integer $l$ marks the lattice sites of the chain. 
The equations of motion read  
\begin{equation} 
\dot{x}_l = p_l\;\;,\;\; \dot{p}_l = -\frac{\partial W_p}{\partial x_l} 
\;\;. 
\label{2-2} 
\end{equation} 
Since we are interested in obtaining solutions which decay to zero 
at spatial infinities and can be interpreted as excitations above 
some regular ground state $x_l=0$, we demand that $h(s)$ behaves regularly 
and especially that $h(s \rightarrow \pm \infty)$ does not diverge. 
The potential energy $W_p$ is a homogeneous function of the 
coordinates $x_l$ since $W_p(\{\lambda x_l\}) = \lambda^2 W_p(\{ x_l\})$. 
The homogeneity of the potential function can be used to separate 
time ($G(t)$) and space ($u_l$) variables as done e.g. in 
\cite{ysk93-pre},\cite{sf94}: 
\begin{equation} 
x_l(t) = u_l G(t)\;\;,\;\;\ddot{G} = -\kappa G\;\;,\;\; 
-\kappa u_l = -\frac{\partial W_p(\{ u_{l'}\})}{\partial u_l} 
\;\;. 
\label{2-3} 
\end{equation} 
Here $\kappa > 0$ is needed to ensure boundness of the solutions. 
Indeed the time dependence is then given by 
\begin{equation} 
G(t) = A \cos (\omega t + \phi)\;\;, \;\; \omega^2 = \kappa\;\;. 
\label{2-4} 
\end{equation} 
The equations for the spatial amplitudes $u_l$ read 
\begin{eqnarray} 
\kappa = \left( 1 - \frac{u_{l+1}}{u_l}\right)  h(s_l) + \left( 1 - 
\frac{u_{l+1}}{u_l} \right)^2 h'(s_l) p_l  \nonumber \\
+ \left( 1 - \frac{u_{l-1}}{u_l} \right) h(s_{l-1}) - 
\left( 1 - \frac{u_{l-1}}{u_l} \right)^2 h'(s_{l-1}) p_{l-1}  
\;\;, \label{2-5} \\
p_l= \frac{u_l}{u_{l+1}} - \frac{u_{l+1}}{u_l} 
\label{2-6} 
\end{eqnarray} 
and $s_l\equiv s_l(\{u_{l'} \})$. 
Here $f'(x)$ means the first derivative of $f$ w.r.t. $x$. 
We are looking for a solution of $\{ u_{l'} \}$ 
which is localized in space, i.e. $u_{l \rightarrow \pm \infty} 
\rightarrow 0$. 
In order to find such a solution to (\ref{2-5}) we 
assume that $s_l$ is constant in the spatial tails. This condition 
is equivalent to having exponential decay.  
Taking 
\begin{equation}
u_l = (-1)^l {\rm e}^{-\beta |l|}
\label{solution}
\end{equation}
and combining the two 
different cases $l=0$ (center of the solutions) and $l \neq 0$
(tails of the solution) we find 
\begin{equation} 
h(-s) = -s(s+1)h'(-s) \;\;,\;\; s=2\cosh (\beta)
\label{2-7}
\end{equation}
and 
\begin{equation}
\kappa = \left(2+s(1-\gamma)\right) h(-s)
\left[ 1+ \frac{\gamma}{s+1}\gamma \left(1+\frac{s}{2}(1-\gamma) \right) \right]
\;\;,\;\; \gamma = \sqrt{1-\left( \frac{2}{s} \right) ^2 }\;\;.
\label{2-8}
\end{equation}
Suppose there exists a value for $s=s_0 > 2$ such that (\ref{2-7})
is satisfied. If for this value we also have $h(-s) > 0$ then
(\ref{2-8}) defines a positive value for $\kappa$. Moreover the
found solution $s_0$ of (\ref{2-7}) will be structurally stable against
changes in $h(s)$. Such models can be generated by starting with
the trivial case $h(s)=1$. A strong enough local distortion in
$h(s)$ at
$s < -2$ will generate solutions of the above equations, preserving
the overall positivity and boundness   $a > h(s) > 0$ for all $s$ with
finite $a$. 
Thus the state
$x_l=c$ (here $c$ is an arbitrary constant) will be a state with minimum energy $H=0$, with all other
trajectories  having larger energies. We can therefore interprete the
found explicit localized time-periodic solution (\ref{solution}) 
as a discrete breather
solution, an excitation above a classical homogeneous ground state.

Let us consider an example. Choose 
\begin{equation}
h(s) = 1 + a {\rm e}^{-bs^2}
\;\;.
\label{2-9}
\end{equation}
For parameters 
\begin{equation}
a > \frac{{\rm e}^{3/2}}{2 + 3 \sqrt{3/(2b)}}
\;\;,\;\; b > \frac{3}{8}
\label{2-10}
\end{equation}
there is at least one solution to (\ref{2-7}) with $s > 2$. 
This solution is structurally stable against perturbations
in $h(s)$. E.g. for $b=0.5$ and $a=0.7$ the solution is $s\approx
2.06781$, $\beta \approx 0.25967$.

Why did we choose a staggered solution $u_l \sim (-1)^l$ ? This
is motivated by the fact that we consider perturbed 
harmonic chains ($h(s)=1$) for which the spectrum of plane waves
is acoustic. Thus discrete breathers in order to localize should
show up with frequencies above the acoustic phonon band which
implies out-of-phase motion of nearest neighbours, or simply staggered
solutions. It is not that simple to apply these arguments to the
case of $h(s) \neq 1$, since there is no simple way to linearize
the equations of motion around $x_l=0$ in the general case.
Without going into further details let us state here, that using
a nonstaggered ansatz one arrives at equations which do not ensure
positivity of $\kappa$ in general. 

Let us study the problem of small amplitude excitations
above the groundstate of (\ref{2-1}). First we recall that we 
consider only positive and bounded functions $h(s)$. Then there
exists a continuous family of ground states, i.e. solutions with
lowest possible energy $E=0$ and $\dot{x}_l=0$ which are given
by 
\begin{equation}
x_l=c\;\;.
\label{2-11}
\end{equation}
Notice that the Hamiltonian (\ref{2-1}) is not invariant under
transformations $x_l \rightarrow c + x_l$, yet the ground state
energy is degenerate. An expansion around one of the ground states
yields 
\begin{equation}
W_p=\frac{c^2}{2} \sum_l \left[ h(2)\left( \delta_l - \delta_{l+1} \right) ^2
+ h'(2) \left( \delta_l - \delta_{l+1} \right) ^4 + O(\delta ^5) \right]
\;\;,\;\; \delta_l = \frac{x_l}{c}-1
\;\;.
\label{2-12}
\end{equation}
The neglected terms of fifth and higher orders in (\ref{2-12})
are not invariant under transformations $\delta _l \rightarrow \tilde{c} 
+ \delta _l$. However the terms up to fourth order are invariant.
Taking into account only terms up to fourth order thus yields
a so-called Fermi-Pasta-Ulam chain for the dynamics of small deviations
from the ground state $x_l=c$. Note that we can not simply perform
the limit $c \rightarrow 0$ since the expansion (\ref{2-12}) is
valid only if $|\delta_l| \ll c$. 
The above found breather solution, which decays to zero at infinities
(and not to $c\neq 0$), can not be easily deformed in order to
decay to $c\neq 0$ at infinities. However it is wellknown that
(\ref{2-12}) allows for discrete breather solutions if $h'(2) > 0$ (which 
can not be given in a closed analytical form) (e.g. 
\cite{st88}, \cite{jbp90}, \cite{sp95}). So we conclude that
for the considered model (\ref{2-1}) discrete breather excitations
above the ground state $x_l=c\neq 0$ are solutions with generic
features, and the ground state $x_l=0$ 
allows
for discrete breather excitations given in a closed analytical form.

In the last part of this section we will discuss the existence
of different variants of discrete breathers with $x_{l \rightarrow \infty}
\rightarrow 0$ asymptotics. The existence of the above
derived discrete breather solution can be interpreted as follows.
Our ansatz $u_l = (-1)^l {\rm e}^{-\beta |l|}$ contains together 
with the parameter $\kappa$ (coming from separating time and space
in (\ref{2-3})) two parameters to be determined - $\beta$ and $\kappa$.
With our ansatz we found two equations - one for the spatial wings
of the solution, and one for the center. Two equations with two variables
can be solved in general, and the additional inequality $\kappa > 0$
will serve as an additional restriction for the choice of possible functions
$h(s)$, but will not change the fact that once solutions are found, they
will be in general structurally stable against changes in $h(s)$. 
Let us now look for a solution of the form $u_l={\rm e}^{-\beta l}$ for
$l \ge 0$, $u_l=a{\rm e}^{\beta ' l}$ for $l \le -1$. 
We now have four equations - two in the tails, one at $l=0$ and
one at $l=-1$, and four parameters - $\kappa, \beta, \beta ', a$. So we
conclude that in general such solutions will exist. Indeed, the solution
from above is a variant of the more general case discussed here with
$a=1$ and $\beta ' = \beta$. 
So we can expect in general a countable set of other solutions
with $a\neq 1$ and $\beta ' \neq \beta$. All these solutions will
have a closed analytical form with parameters to be determined numerically
from the mentioned equations.

\section{Site ordered models in $d$ dimensions}

Consider a $d$-dimensional hypercube lattice with a scalar
coordinate $x_{\bf l}$ associated to each lattice site. The site
index ${\bf l}=(l_1,l_2,...,l_d)$ is a $d$-dimensional vector with integer
components $l_i$.
Consider the operator $\hat{L}$ defined as
\begin{equation}
\hat{L} x_{\bf l} = \sum_{{\bf l'}, |{\bf l'} - {\bf l}|=1}x_{\bf l'}
\;\;.
\label{3-1}
\end{equation}
Here $|{\bf l}|^2 = l_1^2 + l_2^2 + ... + l_d^2$. 
Also define 
\begin{equation}
s_{\bf l} = \frac{\hat{L} x_{\bf l}}{x_{\bf l}}
\;\;.
\label{3-2}
\end{equation}
The Hamiltonian is given by a sum over kinetic and potential
energies as in (\ref{2-1}), with the potential energy  
\begin{equation}
W_p = \frac{1}{2}\sum_{\bf l}\left(\hat{L} x_{\bf l} \right)^{2}
h(s_{\bf l})
\;\;.
\label{3-3}
\end{equation}
Assuming time-space separability as in (\ref{2-3})
we obtain
\begin{equation}
\ddot{G} = -\kappa G
\label{3-4}
\end{equation}
for the time dependence. Again we need $\kappa > 0$ to ensure
bounded solutions. 

The spatial coordinates have to satisfy an equation similar to
(\ref{2-3}). Let us calculate the derivative
\begin{equation}
\frac{\partial W_p}{\partial u_{\bf l}} =
\frac{\partial }{\partial u_{\bf l}} \left[ \frac{1}{2}
\left(\hat{L} x_{\bf l} \right)^{2} h(s_{\bf l}) \right]
+
\frac{\partial }{\partial u_{\bf l}} \hat{L}
\left[ \frac{1}{2} \left(\hat{L} x_{\bf l} \right)^{2} h(s_{\bf l}) \right]
\;\;.
\label{3-5}
\end{equation}
Evaluation of these equations leads to the result
\begin{equation}
\kappa u_{\bf l} = -\frac{1}{2} u_{\bf l} s_{\bf l}^3 h'(s_{\bf l})
+ \hat{L} \left[ u_{\bf l} s_{\bf l} h(s_{\bf l}) \right]
+ \frac{1}{2} \hat{L} \left[ u_{\bf l} s_{\bf l}^2 h'(s_{\bf l}) \right]
\;\;.
\label{3-6}
\end{equation}
A closer study of equation (\ref{3-6}) shows that it could be easily
solved if $s_{\bf l}$ is essentially constant everywhere on the
lattice, more precisely everywhere except for one site ${\bf l} = {\bf 0}$:
\begin{equation}
s_{{\bf l} \neq {\bf 0}} = s_1 \;\;, \;\;
s_{{\bf l} = {\bf 0}} = s_0\;\;.
\label{3-7}
\end{equation}
In that case evaluation of (\ref{3-6}) yields
\begin{eqnarray}
\kappa = s_1^2 h(s_1)\;\;,\;\; |{\bf l}| > 1\;\;,
\label{3-8} \\
\kappa = s_1^2 h(s_1) +\frac{u_{\bf0}}{u_{\bf l}}
\left[ s_0h(s_0) + \frac{1}{2}s_0^2h'(s_0) -
s_1h(s_1) - \frac{1}{2}s_1^2h'(s_1)\right]
\;\;,\;\;|{\bf l}|=1\;\;,
\label{3-9} \\
\kappa = s_0^2h(s_0)\;\;,\;\;{\bf l} = {\bf 0} \;\;.
\label{3-10}
\end{eqnarray}
Defining a new function
\begin{equation}
g(s)=\frac{1}{2}s^2h(s)
\label{3-11}
\end{equation}
equations (\ref{3-8})-(\ref{3-10}) are reduced to
\begin{equation}
\kappa = 2g(s_0)\;\;,\;\;g(s_0)=g(s_1)\;\;,\;\;
g'(s_0)=g'(s_1)\;\;.
\label{3-12}
\end{equation}
These relations (\ref{3-12}) are in fact conditions on the
choice of Hamiltonians, i.e. of the function $h(s)$. 
They are so far rather general, but we have to specify
the values $s_0,s_1$. These values will be connected with
the solution of (\ref{3-6}) through the conditions (\ref{3-7}),
which actually constitute a linear equation:
\begin{equation}
\hat{L} u_{\bf l} = s_{\bf l} u_{\bf l}\;\;.
\label{3-13}
\end{equation}
A solution to this equation can be cast into the form
\begin{equation}
u_{\bf l} = \int{\rm d}q_1 {\rm d}q_2...{\rm d}q_d
\frac{\cos (l_1q_1) \cos (l_2q_2)... \cos (l_dq_d)}{
2\left(\cos (q_1) + \cos (q_2) + ... \cos (q_d)\right) -s_1}
\;\;,
\label{3-14}
\end{equation}
where the integration extends for each variable $q_i$ from $-\pi$
to $\pi$. Then the value for $s_0$ is given by
\begin{equation}
\frac{1}{s_0 - s_1} = \int{\rm d}q_1 {\rm d}q_2...{\rm d}q_d
\frac{1}{2\left(\cos (q_1) + \cos (q_2) + ... \cos (q_d)\right) -s_1}
\;\;.
\label{3-15}
\end{equation}
In other words, given the solution (\ref{3-14}) to (\ref{3-6}), we
can generate the corresponding Hamiltonian by solving (\ref{3-12})
with the additional constraint (\ref{3-15}), which fixes the value
$s_0$ relative to $s_1$. The function $g(s)$ can then be constructed
in the following way: first choose $s_1$, then determine $s_0$,
finally find a function $g(s)$ which is positive in 
$s_1$ and whose values and derivatives are equal in both $s_0$ and
$s_1$. This function $g(s)$ will then define $h(s)$ with (\ref{3-11})
and thus the Hamiltonian with potential energy (\ref{3-3}).

Notice that (\ref{3-14}) is exponentially localized around ${\bf l} = {\bf 0}$
if $|s_1| > 2d$. 

\section{Conclusions}

We obtained explicit discrete breather solutions for several
classes of Hamiltonian lattice models with different lattice dimensions. 
The outlined approach can be extended to other classes of
lattice models with homogeneous potentials. 

An interesting question for future studies will be the 
linear stability analysis of the obtained solutions, which
is also closely related to the spectrum and character of
small amplitude fluctuations around the state $x_l=0$. Note
that any solution which satisfies time-space separability can
be continuously tuned to $x_l=0$ by letting $A \rightarrow 0$ in
(\ref{2-4}), regardless of its spatial profile.

Finally one can extend the solutions of the site-ordered
models to multisite breathers, for which 
\begin{equation}
s_{{\bf l}\neq {\bf l}_m} = s_{\infty}\;\;,
\;\;s_{{\bf l}_m} = s_m
\label{4-1}
\end{equation}
where the number of 'defect' sites ${\bf l}_m$ is finite and
the largest distance between any pair of 'defect' sites is
finite. Work is in progress.
\\
\\
$^1$ Permanent address: Institute of Chemical Physics, Kosygin St. 4,
117334 Moscow, Russia.

\end{document}